\documentclass[twocolumn,showpacs,superscriptaddress]{revtex4-1}
\usepackage{graphicx,amsmath,amssymb,amsfonts,appendix}
\usepackage[colorlinks,linkcolor=blue,citecolor=blue,urlcolor=blue]{hyperref}
\usepackage{color}

\begin{document}
\title{Community structure benefits the fixation of cooperation under strong selection}
\author{Zhi-Xi Wu}\email{eric0724@gmail.com}
\affiliation{Institute of Computational Physics and Complex Systems, Lanzhou University, Lanzhou, Gansu 730000, China}
\author{Zhihai Rong}\email{zhihai.rong@gmail.com}
\affiliation{Web Sciences Center, University of Electronic Science and Technology of China, Chengdu Sichuan 611731, China}
\affiliation{Department of Electronic and Information Engineering, The Hong Kong Polytechnic University, Kowloon, Hong Kong}
\author{Han-Xin Yang}
\email[]{hxyang01@gmail.com}
\affiliation{Department of Physics, Fuzhou University, Fuzhou 350108, China}

\date{Received: date / Revised version: date}
\begin{abstract}
Recent empirical studies suggest that heavy-tailed distributions of human activities are universal in real social dynamics [Muchnik, \emph{et al.}, Sci. Rep. \textbf{3}, 1783 (2013)]. On the other hand, community structure is ubiquitous in biological and social networks [M.~E.~J. Newman, Nat. Phys. \textbf{8}, 25 (2012)]. Motivated by these facts, we here consider the evolutionary Prisoner's dilemma game taking place on top of a real social network to investigate how the community structure and the heterogeneity in activity of individuals affect the evolution of cooperation. In particular, we account for a variation of the birth-death process (which can also be regarded as a proportional imitation rule from social point of view) for the strategy updating under both weak- and strong-selection (meaning the payoffs harvested from games contribute either slightly or heavily to the individuals' performance). By implementing comparative studies, where the players are selected either randomly or in terms of their actual activities to playing games with their immediate neighbors, we figure out that heterogeneous activity benefits the emergence of collective cooperation in harsh environment (the action for cooperation is costly) under strong selection, while it impairs the formation of altruism under weak selection. Moreover, we find that the abundance of communities in the social network can evidently foster the fixation of cooperation under strong-selection, in contrast to the games evolving on the randomized counterparts. Our results are therefore helpful for us to better understand the evolution of cooperation in real social systems.

\end{abstract}
\maketitle

\section{Introduction}
Cooperation means that one individual pays a cost for another to receive a benefit, which is essential for constructing new levels of organization in biology and society~\cite{Nowak2006book,Sigmund2010book}. The emergence of genomes, cells, multi-cellular organisms and human society are all based on cooperation~\cite{Smith1995book}. In fact, ``natural cooperation" has been proposed to be regarded as a third fundamental principle of evolution beside mutation and natural selection~\cite{Nowak2006science}. Evolutionary game theory is the powerful mathematical framework for modeling evolution in biological, social, and economical systems~\cite{Smith1982book,Webull1995book,Hofbauer1998book}. In the standard setup of evolutionary game theory, strategies available for the game are represented by a fraction of individuals in the population. Individuals then collect payoffs according to the rules of the game, which will be measured in terms of reproductive fitness. Reproduction can be either genetic or cultural. In the
latter case (usually describing social situation), the strategy of someone who does well is imitated by others.

The most prominent metaphor to study the evolution of cooperation is given by the Prisoner's dilemma (PD)~\cite{Axelrod2006book, Ohtsuki2006nature}: in pairwise interactions, cooperation provides a benefit $b$ to the partner at some cost $c$ to the cooperator ($b>c$), while defection neither bears any costs nor provides any benefits. In an unstructured population, where all individuals are equally likely to interact with each other, defectors have a higher average payoff than unconditional cooperators. Therefore, natural selection increases the relative abundance of defectors and drives cooperators to extinction.

Only in the presence of additional incentive mechanisms can cooperative behavior that increases the fitness of others at a cost to oneself be promoted by natural selection. One of the mechanisms is based on population structure~\cite{Nowak1992nature}, which can lead to clustering of cooperating agents (or social viscosity). Such mechanism is also explained as spatial (or network) reciprocity~\cite{Nowak2006book}. In the past decade, with the booming development of network science~\cite{Barrat2008book,Newman2010book}, much attention has been devoted to evolutionary games on complex population structures characterized by complex networks, where the nodes represent individuals and links represent social relationships~\cite{Szabo2007pr,Roca2009review,Perc2001bio}. Many topological characteristics of complex networks, such as the power-law distribution of interaction degree~\cite{Santos2005prl}, the clustering coefficient~\cite{Szabo2005pre,Vukov2006pre,Wu2005pre, Assenza2008pre,Rong2010pre}, degree-degree correlation~\cite{Rong2007pre}, are proven to have great effects on the evolutionary success of cooperation.

\begin{figure}
\includegraphics[width=0.49\linewidth]{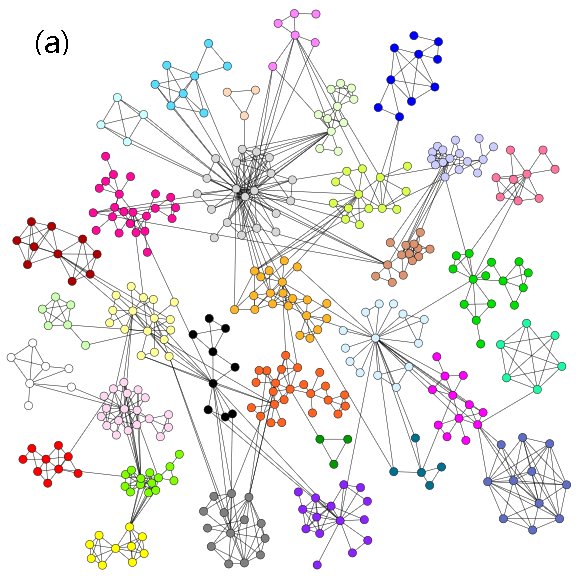}
\includegraphics[width=0.49\linewidth]{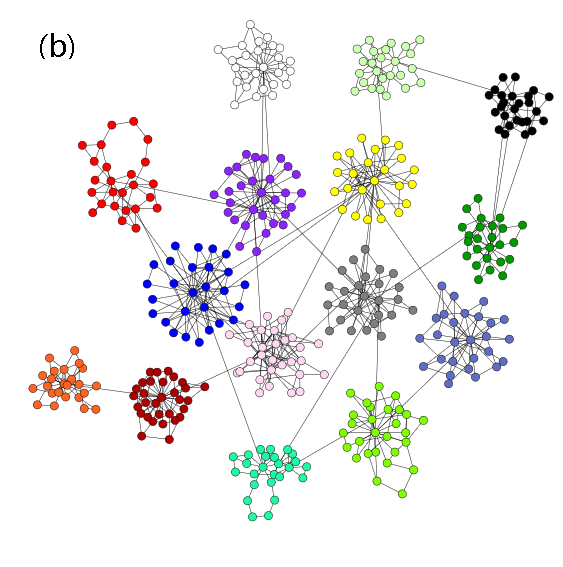}
\includegraphics[width=0.99\linewidth]{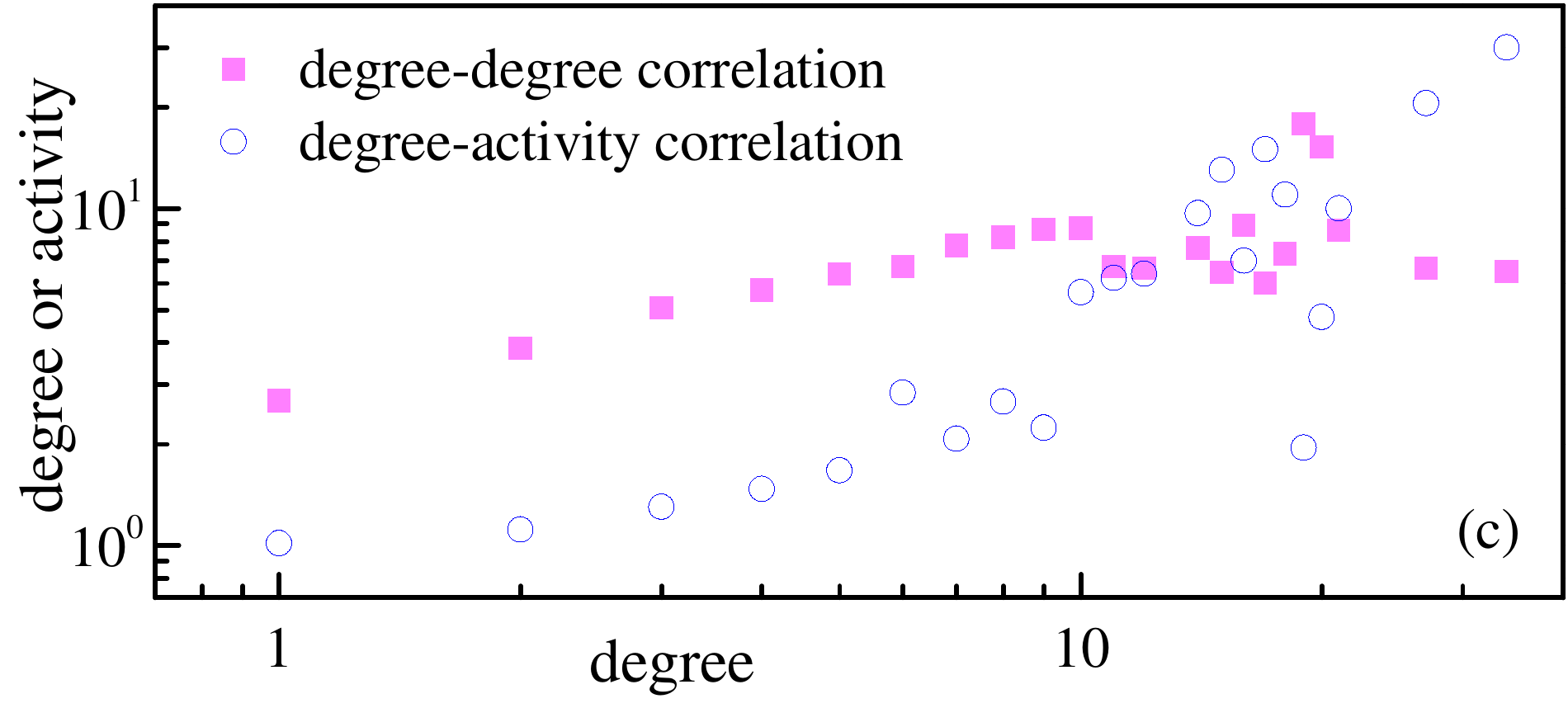}
\includegraphics[width=\linewidth]{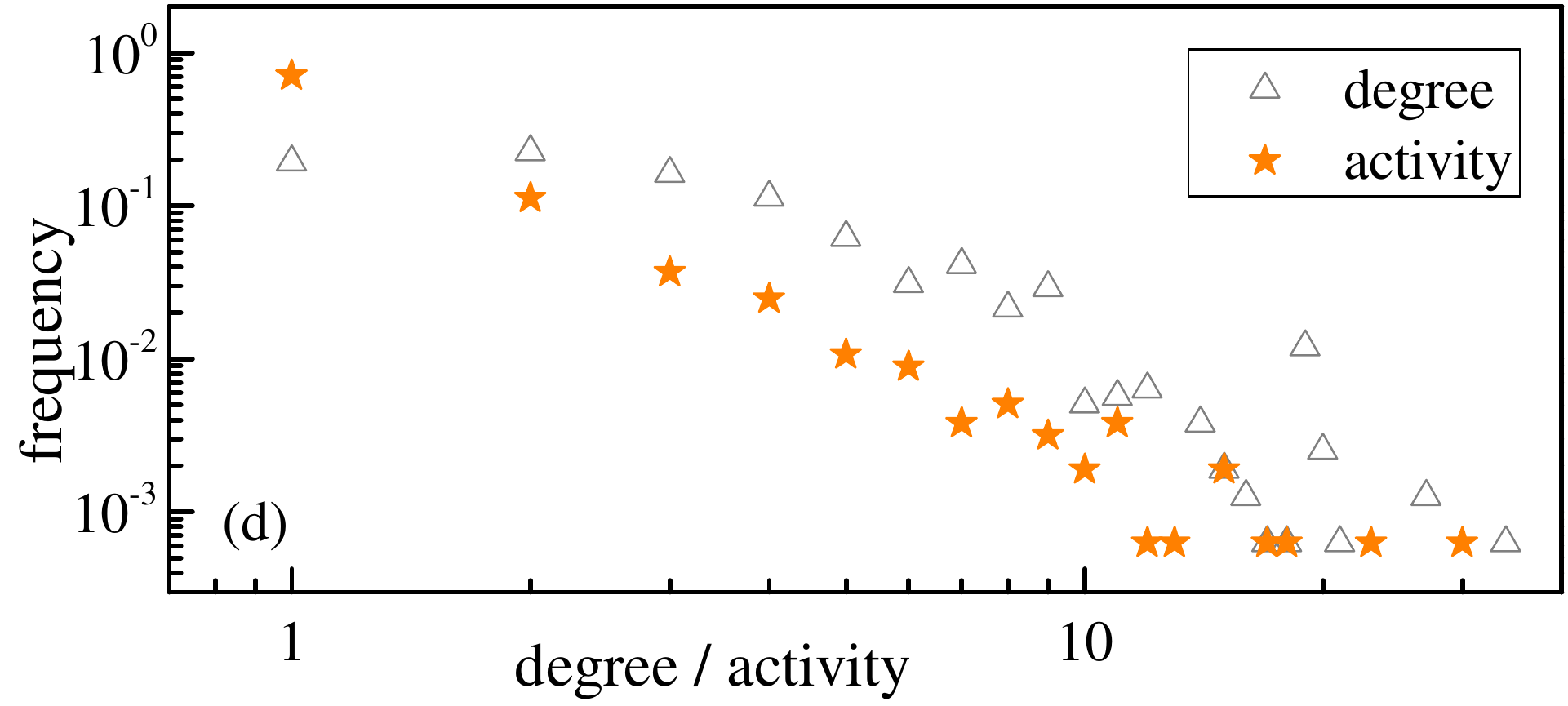}\\
\caption{(Color online) Graphical illustration of (a) the collaboration network of network scientists~\cite{Newman2006pre} (note that only the largest connected component of the network is shown, which contains $N=379$ nodes and $M=914$ links), and of (b) a synthetic community network with 15 distinct communities (comprising of $N=400$ nodes and $M=858$ links), generated by means of the method proposed in~\cite{Lancichinetti2008pre}. The nodes with different colors belong to distinct communities, as revealed using the community detection algorithm VOS Clustering~\cite{vos}. (c) The average activities and mean degree of nearest neighbors as a function of the nodes' interaction degree display clear positive (or assortative) correlations. (d) Heavy-tailed distributions of the node degree and activity of the collaboration network. Note that the measures in (c) and (d) are calculated for the whole collaboration network containing $N=1589$ noes and $M=2742$ links.}\label{network}
\end{figure}

One of the most common and important features of a social network is community structure~\cite{Newman2001pnas,Newman2006pnas, Newman2012np}, the division of network nodes into groups within which the network connections are dense, but between which they are sparser [see Fig.~\ref{network}(a) for a typical real social network]. The strength of the community structure can be quantified by using a modularity measure~\cite{Newman2004pre}, which for a random network will be close to zero, and will be close to one for a strong community structure. Though community structure is ubiquitous in social and biological networks, it is not enough clear until now whether the community structure promotes or inhibits cooperation. Evolutionary PD game on complex networks with community structure have been studied only rarely. In Ref.~\cite{Chen2006pa}, Chen \emph{et al.} studied the PD on community networks, but focused mainly on how varying the inter-community links affects cooperation. Along similar lines, recent work by Lozano \emph{et al.}~\cite{Lozano2008plos} suggests the inter- and intra-structure of the communities influence the evolution of cooperation in
social networks by raising or lowering the level of cooperation and
the stability of the behavior of the communities against changes on
the temptation to defect. More recently, in Ref.~\cite{Wu2014pa}, the authors studied the PD game on community network and Erd\H{o}s-R\'{e}nyi (ER) random graphs, and found that community structure usually inhibits cooperation. We note that the community network in~\cite{Wu2014pa} is actually composed of two ER random graphs with a few links connecting them, which shouldn't be regarded as a good candidate for real social systems.

On the other hand, recent empirical studies suggested that heavy-tailed distributions (usually of power-law form) of human activities are universal in real social dynamics~\cite{Radicchi2009pre,Iribarren2009prl, Cattuto2010plos,Muchnik2013sr}, which means that a significant number of individuals are much more active than most others. Inspired by these two things, i.e., evident community structure and heterogeneity in individual activity in social networks, we in this work intend to study evolutionary PD games on a real social network with distinct communities and with heterogeneous individual activity. Our main purpose is to figure out how such typical ingredients of real social systems affect the evolution of cooperation.

\section{Model}
Let us first introduce our networked game model. We will focus our attention on evolutionary PD games on the collaboration network of network scientists~\cite{Newman2006pre}. The largest connected component in this collaboration network contains $N=397$ nodes with $M=914$ connections among them (see Fig.~\ref{network}). The strength $s_{ij}$ of the social tie (or contact weight) between two nodes $i$ and $j$ can be reasonably formulated as described in~\cite{Newman2001pre}
\begin{equation}\label{strength}
s_{ij}=\sum_k\frac{\delta_{i}^{k}\delta_{j}^{k}}{n_k-1},
\end{equation}
where $\delta_{i}^{k}$ in the numerator is $1$ if scientist $i$ was a coauthor of paper $k$ and zero otherwise, and $n_k$ in the denominator is the number of coauthors of paper $k$ involving scientists $i$ and $j$. The activity of individual $i$ is therefore conveniently defined by summing the contact weight over all adjacent neighbors as $A_i=\sum_{j\in\Omega_i}s_{ij}$. Note that $A_i$ is no other than the total number of papers that individual $i$ has coauthored with others.

Each player $i$ occupies a vertex of the collaboration network, and engages in pairwise interactions to collect a payoff, $P_i$, from game interactions within its neighbors, as specified by the underlying interaction network. Initially, all the individuals can choose either to cooperate or to defect with equal probability. A cooperator pays a cost, $c$, for each neighbor to receive a benefit, $b$. A defector pays no cost and provides no benefit. The payoffs accrued in the game interactions by the players will influence their reproductive fitness (or the propensity to propagate their strategies). For simplicity, we adopt a rescaled payoff matrix such that there is only one single parameter for the PD with all the original properties retained\\ \\
\centerline{
\begin{tabular}{r|c c}
&$C$ &$D$ \\
\hline
$C$ &$1$ &$0$ \\
$D$ &$1+c/b$ &$c/b$,\\
\end{tabular}
}\\ \\
where $c/b$ is the cost-to-benefit ratio of cooperative behavior~\cite{Nowak2006book}. Following~\cite{Ohtsuki2006nature}, we assume that the fitness (or performance) of an individual is given by a constant term, denoting the baseline fitness, plus the payoff that arises from the game. In particular, the fitness of a player $i$ is given by $f_i=1-w+wP_i$, where $w$ weighs how the payoffs acquired from the game interactions contribute to the fitness, thus $w$ measures the intensity of selection. Strong selection and weak selection correspond, respectively, to the cases of $w\rightarrow 1$ and $w\ll 1$, which describe the situations that the payoff is either large or small compared to the baseline fitness. The idea behind weak (strong) selection is that many different factors contribute to the overall fitness of an individual, and the game under consideration is just one (the most important one) of those factors~\cite{Ohtsuki2006nature}.

The evolutionary process is preceded by implementing a variation of birth-death process for the population, comprising the following elementary steps. I) fitness accumulation process. In this stage, the players are allowed to interact with their neighbors with certain probabilities to collect profits, which will constitute their evolutionary fitness. To take into account explicitly the heterogeneous activity of individuals, we select players proportional to their activities to take participating in game interactions. For the sake of comparison, we also consider the case that at each time step all the individuals are chosen randomly to playing games. In both cases, one Monte Carlo step (MCS) consists of all the players having interacted with their neighbors once on average. II) Birth-death (or reproduction) process. After every MCS, each player reproduces $k_i+1$ (where $k_i$ is the interaction degree of $i$) offsprings, and $k_i$ of them are dispersed to those immediate neighboring sites and one left in the original site. Subsequently, all the \emph{old} individuals are chosen to die, and the \emph{new} born offsprings will compete for the empty sites in proportion to their fitness, and only one individual is allowed to survive on each site after the competition. Note that the number of competing individuals on site $i$ is still $k_i+1$. Thus, the probability of a cooperator wining on site $i$ reads as
\begin{equation}\label{winrule}
\frac{F_C}{F_C+F_D}=\frac{\sum_{j\in\Omega_i(C)}f_j} {\sum_{j\in\Omega_i(C)}f_j+\sum_{k\in\Omega_i(D)}f_k},
\end{equation}
where $\Omega_i(C)$ [$\Omega_i(D)$] denotes the set of neighboring cooperators (defectors) of $i$ (including itself) before reproduction. From the social point of view, the considered birth-death process can be reasonably regarded as a ``proportional imitation" rule, where the strategies with better performance in ones neighborhood are more readily to be imitated by the focal individual in the subsequent round.

Since in finite populations evolution is stochastic so as that the combination of selection and random drift eventually leads to fixation of one of the strategies, we iterate the above discrete-time dynamics for a large number of MCS, until the system gets into one of the two absorbing state: fixation of cooperation (composed of only cooperators) or fixation of defection (composed of only defectors). Our key quantity is therefore the average fixation probability of cooperators, which is determined by the fraction of runs where cooperators reached fixation out of $10^5$ independent runs.

\section{Results and analysis}
It is convenient for us to use some abbreviations for the following discussions. Hereinafter, ``WeakRan" and ``WeakAct" (``StrongRan" and ``StrongAct") mean that the game dynamics is under weak-selection (strong-selection) and the individuals are picked randomly or in terms of their actual activities to playing PD games, respectively.

\begin{figure}
\includegraphics[width=\linewidth]{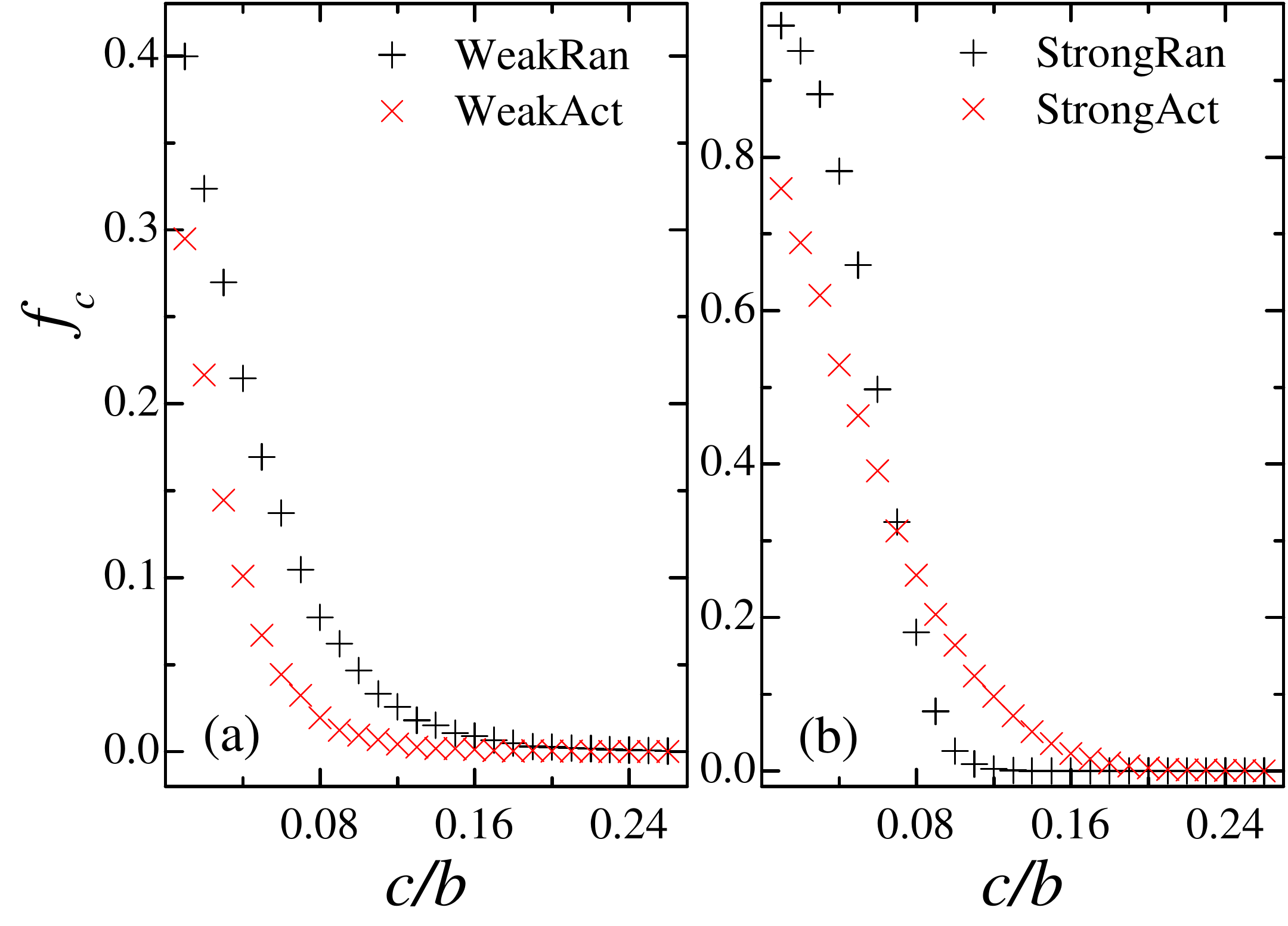}\\
\caption{(Color online) Average number of independent runs ending in all cooperators, i.e., the fixation probability of cooperation $f_c$, as a function of the cost-to-benefit ratio $c/b$ for the PD game taking place on the collaboration network of network scientists~\cite{Newman2006pre}. (a) and (b) correspond to the cases of weak-selection $w=0.01$, and strong-selection $w=1.0$, respectively. The pluses and crosses in each panel are for the results where the individuals are selected at random (Ran) or in terms of their activity (Act) to playing PD games with the adjacent neighbors. The data are yielded by implementing $10^5$ different realizations.}\label{netscience}
\end{figure}

Let us first explore how cooperation evolves on the collaboration network under weak-selection. The average fixation probability of cooperation $f_c$ as a function of the cost-to-benefit ratio $c/b$ in $10^5$ independent realizations is featured in Fig.~\ref{netscience}(a) for both cases of WeakRan and WeakAct. As expected, the smaller the value of $c/b$ is, the less costly being a cooperator is, and thus the larger the probability of fixation to cooperation on the system is. As $c/b$ increases, defection becomes more and more preferable, and the fixation probability of cooperation is vanishing. From Fig.~\ref{netscience}(a), we see that under weak selection, $f_c$ in the case of WeakRan is greater than that in the case of WeakAct, suggesting that the heterogeneous activity of individuals does harm to the emergence of cooperation in the entire range $c/b>0$.

For weak-selection, we have adopted $w=0.01$ in calculating the fitness of individuals, i.e., the payoffs reaped from game interactions contribute very little to the reproductive capability. Consequently, the competition between different strategies behaves mainly like the \emph{random drift} process. In such situation, if the players are picked in terms of their activities to playing games in each MCS, only a few players have the chance to harvest gains (suppose the extreme case where only the player with highest activity is always chosen). As such, during the reproduction process, many sites will be occupied randomly by cooperators and defectors, owing to the fact that nearly no payoff is gained from game interactions (and remind that the baseline fitness for all players is the same), which is definitely detrimental for the dispersal of cooperation. In contrast, if the players are selected randomly to playing games, cooperative behavior could be encouraged since for small $c/b$ mutual cooperation produces the highest group profits and more chance to participating in game interactions would promote the competing capability of cooperators (especially those clustered cooperators).

In fact, as will be shown below in Figs.~\ref{randomized} and~\ref{synthetic}, under weak-selection, any constraint on strategy diffusion, either by the presence of communities in the underlying interaction network or choosing players to playing games in terms of their activities, will diminish the chance for cooperators to expand in the population, hance impair the emergence of cooperation.

Next, we report our simulation results on collaboration network with strong-selection. As shown in Fig.~\ref{netscience}(b), we observe a crossover behavior of $f_c$ as a function of $c/b$ under StrongRan and StrongAct. Specifically, when the value of $c/b$ is sufficiently small, the average fixation probability of cooperators under StrongRan is greater than that under StrongAct, similar to as found in the case of weak-selection. However, as $c/b$ goes beyond a certain value, say $c/b\approx0.078$, we get the reverse result that cooperation is more favored under StrongAct than under StrongRan. We argue that the presence of community structure in the collaboration network contributes to the crossover behavior.

Under strong-selection ($w=1.0$), the payoffs gathered from game interactions will determine totally the reproduction success of the strategies. For sufficiently small $c/b$, cooperative behavior is quite cheaper and the benefit for unilateral defection is very slim. Mutual cooperation will enable the involved cooperators to gain a sustainable competitive advantage. As a consequence, more chance to take part in games means more opportunities for cooperators to take over defecters. Accordingly, for small $c/b$, selection of players to playing games with equal chance is most likely to support cooperation, in comparison to any other biased selection manners.

For sufficiently large $c/b$, cooperative behavior becomes very expensive and will be doomed to extinction. For \emph{large} value of $c/b$ where cooperators have certain, yet small, probabilities to taking over the whole population, say $0.08<c/b<0.16$ in Fig.~\ref{netscience}(b), how the players participate in game interactions will affect the fate of cooperation. For instance, for large $c/b$, a few communities composed of players with large activities may occasionally dominated by cooperators. Once such situation is realized, it will become fairly difficult for these communities to be invaded by defectors, given that the players are selected to playing games in terms of their activities (i.e., the StrongAct case), since the players in these communities are more likely chosen to collecting payoffs through game interactions [note that the degree-activity and degree-degree correlations are both positive (see Fig.~\ref{network}), and remind that connections within communities are very dense, but between which connections are quite sparser]. Thus, as compared to the case of StrongRan, cooperators have a more promising future in the case of StrongAct for large $c/b$.

\begin{figure}
\includegraphics[width=\linewidth]{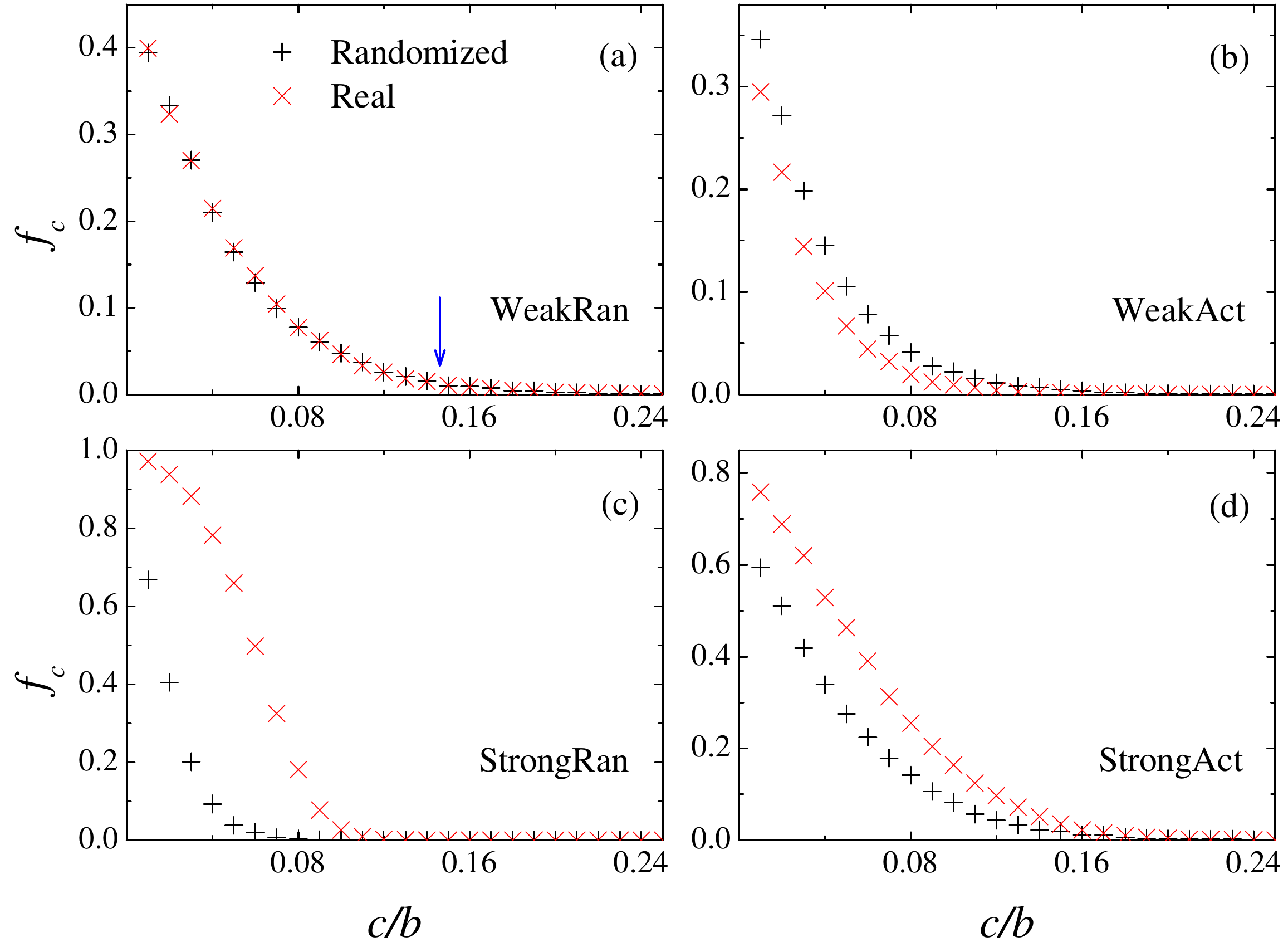}\\
\caption{Average number of independent runs ending in all cooperators, i.e., the fixation probability of cooperation $f_c$, as a function of the cost-to-benefit ratio $c/b$ for the PD game taking place on the real collaboration network as shown in Fig.~\ref{network}(a) (crosses), or on its randomized counterparts (pluses). (a) and (b) are for the weak-selection case $w=0.01$, and (c) and (d) for the strong-selection case $w=1.0$, respectively. The arrow marks the position $c/b=1/(\bar{k}+2)\simeq 0.147$ (here $\bar{k}=914\times 2/379\simeq 4.823$), predicted by Eq.~(\ref{winrule}), below which the fixation probability of cooperation should be $1$ given that the PD game is running on a random regular graph of infinite size. Others are the same as in Fig.~\ref{netscience}.}\label{randomized}
\end{figure}

So far, we have analyzed how the heterogeneity in individuals' activity impacts the evolution of cooperation. We evidenced that under weak-selection, heterogeneous activity always plays a detrimental role in the emergence of cooperation (since it impedes the dispersal of cooperative behavior), while under strong-selection, it is helpful for the surviving and expanding of cooperation when cooperative behavior is costly. The positive role of heterogeneous activity in sustaining cooperation for large $c/b$ is partly due to the community structure present in the underlying social interaction network.

In order to better understand how community structure influences the evolution of cooperation, we next make comparative studies of our current game model on the collaboration network and on its randomized counterparts. We used the method proposed in~\cite{Maslov2004science} to generate randomized versions of the collaboration network. Specifically, the edges connecting different nodes in the original network are repeatedly exchanged such that the distinct community structures are expunged as much as possible, while keeping the degree of each vertex unchanged. No self-connections and isolated networks are allowed in the edge exchange process.

In Fig.~\ref{randomized} we show the average fixation probability $f_c$ of cooperators in randomized collaboration networks as a function of the cost-to-benefit ratio $c/b$ for different game dynamics. The data shown in Fig.~\ref{netscience} are also replotted for comparison. In the case of WeakRan, we observe that the topological structure of the interaction network does not affect much the evolutionary outcome. The symbols for the two networks nearly coincide with each other. This is somewhat expected, since as mentioned before, in the limit of weak-selection, the evolution of cooperation is mainly driven by random drift, and the structure of the underlying interaction network plays as a less important role in determining the fate of cooperation when all the players are chosen with equal chance to playing games. From Fig.~\ref{randomized}(b), we perceive that under weak-selection, if the players are selected to participate in games in terms of their activities, the presence of communities in the underlying interaction network actually undermines the advantages of cooperators, owing to the fact that cooperation could be easily locked in the communities, in contrast to the case of randomized network, where cooperative behavior has a better opportunity to spread out.

\begin{figure}
\includegraphics[width=\linewidth]{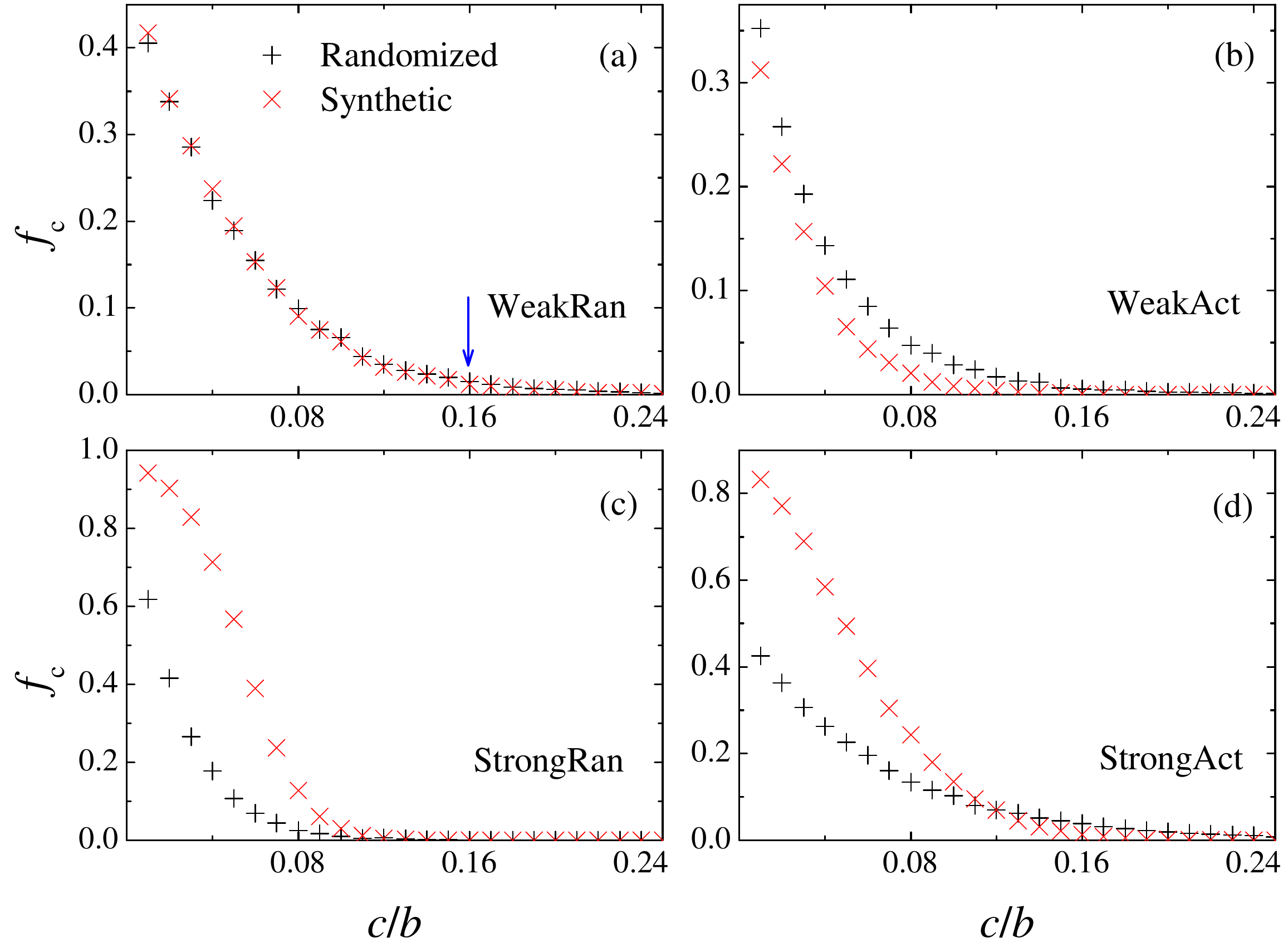}\\
\caption{(Color online) Average number of independent runs ending in all cooperators, i.e., the fixation probability of cooperation $f_c$, as a function of the cost-to-benefit ratio $c/b$ for the PD game taking place on the synthetic community network as shown in Fig.~\ref{network}(b) (crosses), or on its randomized equivalents (pluses). The arrow marks the position $c/b=1/(\bar{k}+2)\simeq 0.159$ (here $\bar{k}=858\times 2/400=4.29$), predicted by Eq.~(\ref{winrule}), below which the fixation probability of cooperation should be $1$ given that the PD game is running on a random regular graph of infinite size. Others are the same as in Fig.~\ref{netscience}.}\label{synthetic}
\end{figure}

The above dynamical scenario reverses if we consider the game dynamics under strong-selection. From Figs.~\ref{randomized}(c) and (d), we clearly see that cooperation is more easily to emerge in social networks with distinct communities than in the randomized counterparts. As explained before, under strong-selection where the payoffs from game interactions matter heavily in determining the reproduction success, the presence of communities in the underlying interaction network (once occupied by cooperators) will provide their members with a stable source of benefits, which ultimately allows them to resist cycles of invasions from defectors. In particular, the promotion of cooperation due to the presence of communities in the real social network under StrongRan [Fig.~\ref{randomized}(c)] is actually not surprising, since in otherwise ``well-mixed case" (games on the randomized network) natural selection always favors defectors over cooperators. Under StrongAct [Fig.~\ref{randomized}(d)], though those more active players would have great chance to get higher profits in the randomized networks if they all adopt cooperation, they are actually more easily to be exploited by defectors because of their \emph{excessive exposure} to others, which in return tends to weaken the expanding of cooperation. If instead they are confined to the communities, whenever mutual cooperation is occasionally emerged among them, the cooperative communities may last for long period of time, which supplies further opportunities for the cooperators to strike back in the competition with defectors and even to dominate the whole population .

It is worthy noting that in the limit of weak selection, by considering the spatial structure of the population, Ohtsuki and Nowak~\cite{Ohtsuki2006jtb} have derived the so-called \emph{spatial} replicator equation for evolutionary PD on a random regular graph of degree $k$, which describes how the average frequency of each strategy changes over time. Remarkably, they found that moving a game from a well-mixed population onto a regular graph simply results in a replicator equation with a transformation of the payoff matrix, whose element is the sum of the original payoff matrix plus another matrix, which describes the local competition of strategies~\cite{Ohtsuki2006jtb,Ohtsuki2008jtb}. For our current PD model with the rescaled payoff matrix and with proportional imitation, if we denote by $x$ the density of cooperators in the population, then its derivative with respect to time in the case of WeakRan should read as~\cite{Ohtsuki2006jtb}
\begin{equation}\label{spatail-replicator}
\dot{x}=x(1-x)\frac{k}{(k+3)(k-2)}[1-(k+2)c/b],
\end{equation}
where $k$ is the degree of the regular graph. Note that Eq.~(\ref{spatail-replicator}) is derived for the game on a graph of infinite size. Thus, cooperators always win over defectors under the condition
\begin{equation}\label{winrule}
1-(k+2)c/b>0\Rightarrow c/b<\frac{1}{k+2}.
\end{equation}
In other words, evolutionary dynamics on random regular graphs can
favor cooperation over defection provided $c/b<1/(k+2)$. In Fig.~\ref{randomized}(a), we also mark the position $c/b=1/(\bar{k}+2)$ [$\bar{k}$ is the average degree of randomized counterparts of the collaboration network in Fig.~\ref{network}(a)], blow (above) which the fixation of cooperation should go to 1 (0). The reason that $f_c$ does not reach 1 is due to the fact that we here consider the PD on a very small system with $N=379$ (actually, even for infinite system size, evolutionary stability on sparse graphs does not imply evolutionary stability in a well-mixed population, nor vice versa~\cite{Ohtsuki2008jtb}), and the randomized counterparts of the collaboration network are not strictly random \emph{regular} graphs at all (because of the heterogeneity in node degree).

Finally, as a further check, let us compare the PD game taking place on a synthetic network with explicit community structures and on its randomized equivalents. The synthetic community network [Fig.~\ref{network}(b)] is generated according to the algorithm proposed in~\cite{Lancichinetti2008pre}. To include heterogeneity in activity, we assign the activity of each node $i$ by the power of its degree as $k_{i}^{1.5}$. The evolutionary results are reported in Fig.~\ref{synthetic}. Once again, we obtain qualitatively similar results as summarized in Fig.~\ref{randomized}, i.e., the presence of communities in the interaction network facilitates the fixation of cooperation under strong-selection, while suppresses it under weak-selection.

\section{Conclusions}
To conclude, we have studied the Prisoner's dilemma game on top of a social network with distinct community structure and on its randomized equivalents. By letting the profits collected by a player after playing with its neighbors to constitute its evolutionary fitness (or performance), we are able to formulate the problem in an evolutionary form. By incorporating realistic ingredients of real social systems into the game dynamics, we have explored how the heterogeneous distribution of individual activities and the presence of communities in the underlying interaction network affect the evolution of cooperation under both weak- and strong-selection. Our main findings are: I) The heterogeneity in individual activities can benefit the emergence of collective cooperation in harsh environment (the action for cooperation is costly) under strong-selection, while it undermines the formation of altruism under weak-selection or under strong-selection provided cooperative behavior is inexpensive; II) The presence of community structure in the underlying interaction network always promotes the evolution of cooperation under strong-selection, while it plays as an opposite role under weak-selection. Since the proposed evolutionary rule~(\ref{winrule}) can be reasonably regarded as a proportional imitation rule, we believe the proposed model is relevant for real social dynamics, and hence the presented results are helpful for us to better understand the evolution of cooperation in our human society, where heterogeneous activities and community structures are general and ubiquitous.

\acknowledgments{}
This work was supported by the National Natural Science Foundation of China (Grant No. 11135001), and by the Fundamental Research Funds for the Central Universities (Grant No. lzujbky-2014-28).

\end{document}